\documentclass[conference]{IEEEtran}
\IEEEoverridecommandlockouts

\usepackage{cite}
\usepackage{amsmath,amssymb,amsfonts}
\usepackage{algorithmic}
\usepackage{graphicx}
\usepackage{textcomp}
\usepackage{xcolor}
\usepackage{scalerel}
\usepackage{tikz}
\usetikzlibrary{svg.path}

\newcommand{\orcidicon}[1]{%
\href{https://orcid.org/#1}{%
\includegraphics[width=1.0em]{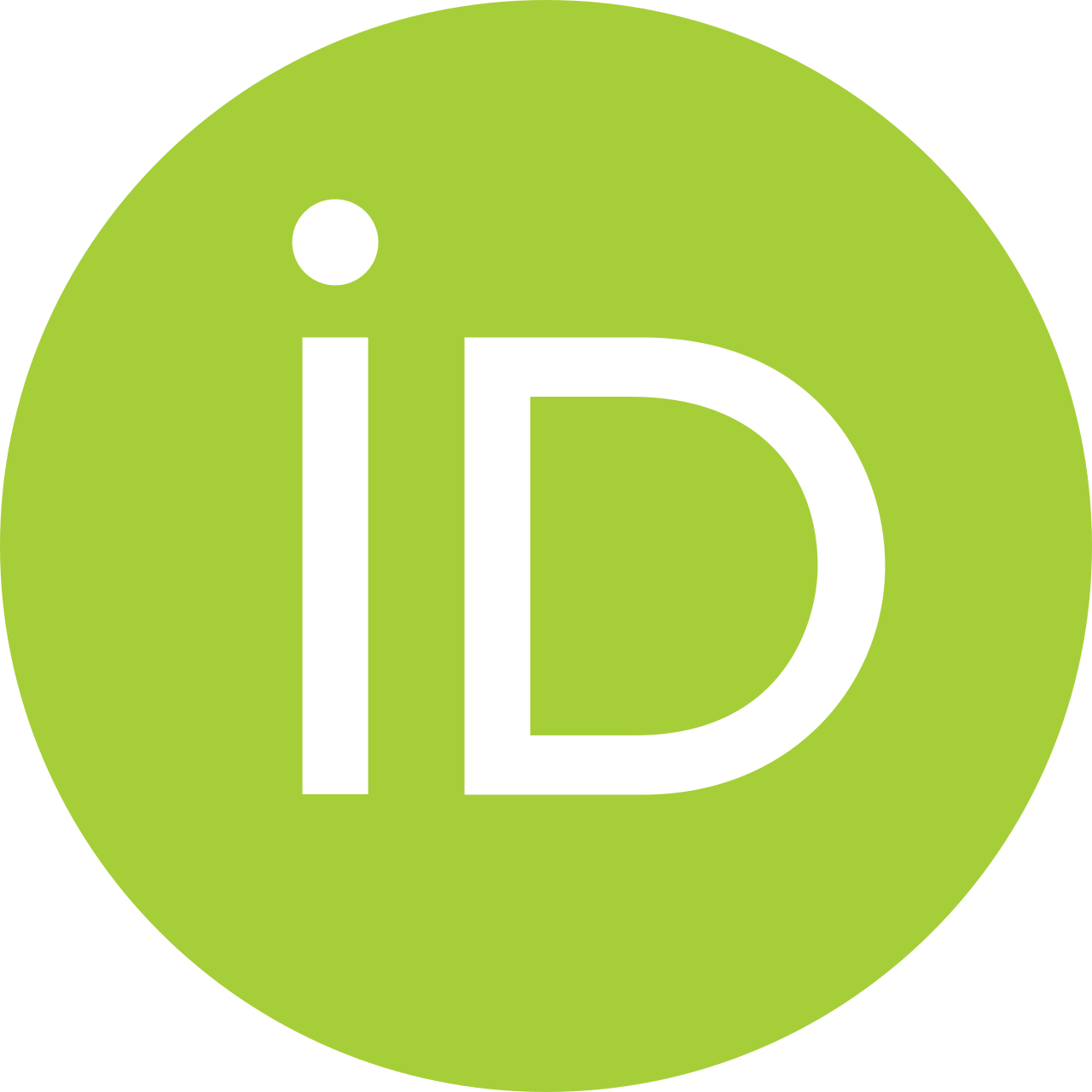}%
}%
}

\usepackage{hyperref} 

\def\BibTeX{{\rm B\kern-.05em{\sc i\kern-.025em b}\kern-.08em
    T\kern-.1667em\lower.7ex\hbox{E}\kern-.125emX}}
\begin{document}

\title{Applied AI-Enhanced RF Interference Rejection\\
\thanks{Research was sponsored by the Department of the Air Force Artificial Intelligence Accelerator and was accomplished under Cooperative Agreement Number FA8750-19-2-1000. The views and conclusions contained in this document are those of the authors and should not be interpreted as representing the official policies, either expressed or implied, of the Department of the Air Force or the U.S. Government. The U.S. Government is authorized to reproduce and distribute reprints for Government purposes notwithstanding any copyright notation herein.}
}

\author{
    Rahul Jain \orcidicon{0009-0009-3723-5720}, Pierre Trepagnier \orcidicon{0000-0003-2869-1504}, Rick Gentile \orcidicon{0009-0004-7758-1947}, Joey Botero \orcidicon{0000-0001-9848-8611}, Alexia Schulz \orcidicon{0000-0002-4143-8792}\\
    \textit{MIT Lincoln Laboratory, Lexington, MA 02421, USA}\\
    \small {\{Rahul.Jain, ptrepagnier, Richard.Gentile, Joey.Botero, Alexia.Schulz\}}@ll.mit.edu\\
}


\maketitle

\begin{abstract}

AI-enhanced interference rejection in radio frequency (RF) transmissions has recently attracted interest because deep learning approaches trained on both the signal of interest (SOI) and the signal mixture (SOI plus interference) can outperform traditional approaches which only consider the SOI. The goal is to detect, demodulate, and decode signals over a range of signal-to-interference-plus-noise (SINR) levels without having a detailed, design-level knowledge of the interfering signal or the propagation conditions. Our present AI interference suppression results are based on Autoregressive Transformer Decoder models which exhibit orders of magnitude faster throughput at inference time than WaveNet models developed in earlier work. As a specific example, we investigate an analog FM ``Walkie Talkie'' radio signal of interest in the presence of an Orthogonal Frequency-Division Multiplexing (OFDM) interferer. This type of interferer is near-ubiquitous in the current RF landscape. Our results clearly show the benefits of transformer-based interference mitigation in tactical settings. We show that unintelligible transmissions become intelligible via metrics such as Perceptual Evaluation of Speech Quality (PESQ), while overall latency is kept to a minimum using readily available lightweight GPUs such as a Jetson AGX Orin. We believe these same techniques can also be applied to a broader set of national security scenarios, as well as having commercial applications.

\end{abstract}


\section{Introduction}
This work reports on AI-enhanced RF interference suppression work undertaken as part of the Department of the Air Force-MIT AI Accelerator (AIA), a collaboration involving the Air Force, MIT, and MIT Lincoln Laboratory. Many of the theoretical advances on the network architectures were originated by our partner researchers at MIT \cite{10945811, jayashankar2025, LifarRFTransformer} and we are primarily focused on operationalizing these advances.

Our research group is generally directed at the use of AI in national security applications, and in particular at the intersection of the cyber domain with the physical---RF systems in this case. In this paper, we examine the use of AI to enhance RF interference rejection  by  radios at the tactical edge. The tactical edge presents a multitude of challenges to the deployment of AI solutions as compared to its usual utilization in academic or enterprise settings. At the tactical edge, an AI deployment is subject to severe size, weight, and power (SWaP) constraints, and yet must perform inference fast enough to be useful. In addition, communication back to enterprise-scale computing capability could be degraded to nonexistent, so the AI's ability to handle modified interferers with zero- or few-shot learning is important\cite{Trepagnier2023}.
Notwithstanding the problems associated with AI at the tactical edge, its advantages for interference rejection are compelling. With the exception of simple repetitive interferers, typically interference mitigation has focused on enhancing the signal of interest (e.g., with a matched filter) rather than suppressing the interferer, due to the latter's non-stationarity. Deep learning based models, however, are trained on the signal mixture as well as the SOI, hence can pick out features of the SOI and interference, enabling interference mitigation. Here, we will demonstrate interference mitigation using AI on analog signals commonly found in tactical radios, while keeping in mind that there are also commercial applications where low complexity and cost-effectiveness are required for mobile and edge devices.

\begin{figure}[htbp]
\centerline{\includegraphics[width=0.75\columnwidth]{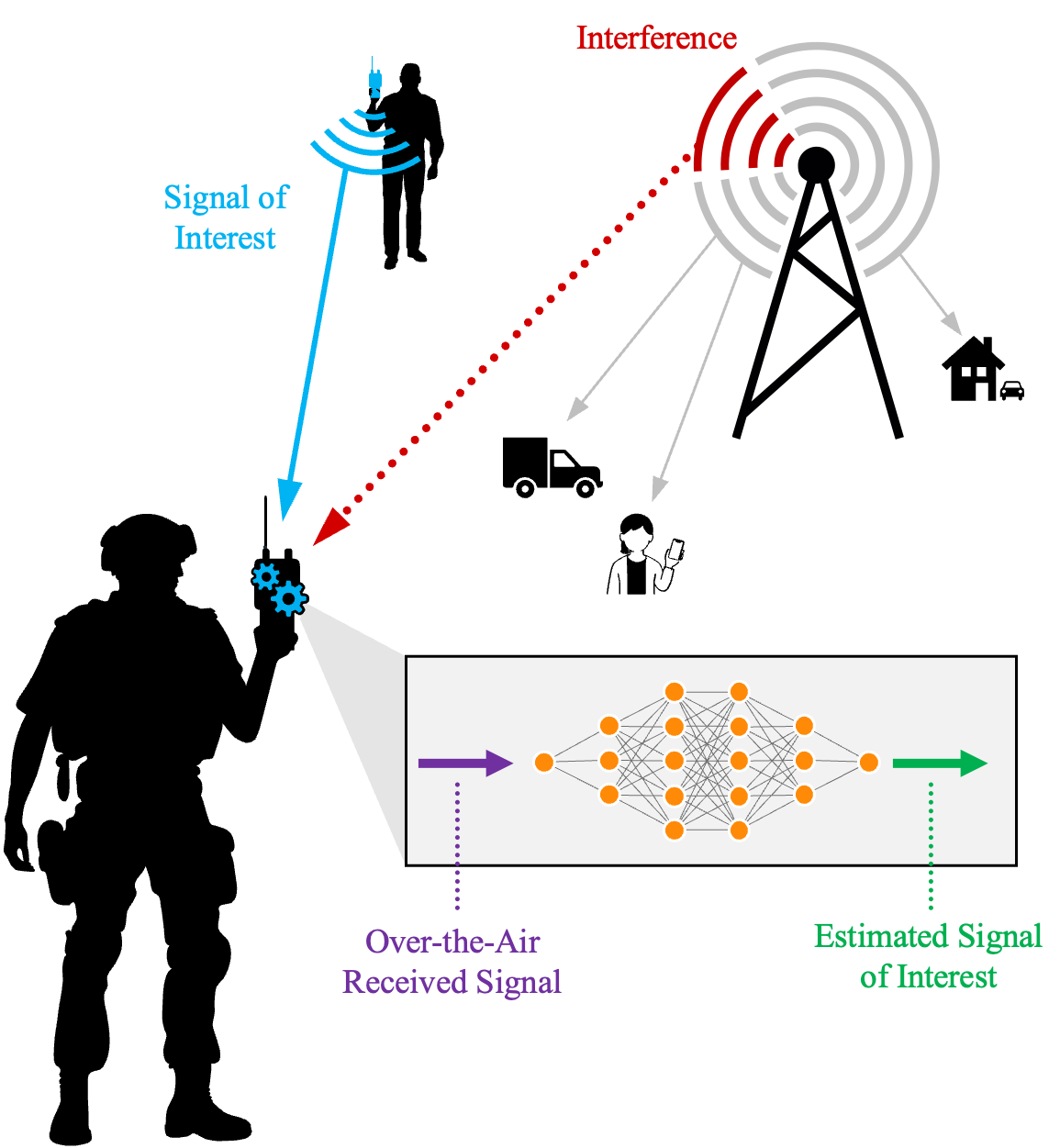}}\caption{Adapted from \cite{10945811}, \cite{11196365}. Example of a co-channel interference scenario for tactical communications where other sources may interfere with a signal of interest and preprocessing is required for recovery.}
\label{fig:problem_formulation}
\end{figure}

In the end application, the algorithm has to work on SOIs over a range of SINR levels. In addition, we will not have a detailed, design-level knowledge of the interfering signal or the propagation conditions. Fig. \ref{fig:problem_formulation} shows our problem setup where we are trying to preprocess a signal mixture to recover a signal of interest for a tactical communications application.

\section{Background}

\subsection{Challenges in the Wireless RF Domain}

Wireless communication systems operate in conditions that can be unpredictable due to challenging radio propagation conditions and presence of intentional and unintentional interference sources. Closing a link budget to recover a signal of interest means that the receiver has to operate on signals over a range of signal-to-interference-plus-noise (SINR) levels. Transmitted signals reach the receiver from multiple paths which introduce different amplitudes, delays, and phases that produce constructive and/or destructive interference at the receiver. Signal fades lead directly to bit errors in the processed signals. 

The crowded RF spectrum also provides the challenge of finding bands where interference is not present. The available frequency bands are rapidly consumed as fast as they become available by commercial applications. In addition, the waveform bandwidth of these systems also grows as commercial software defined radios with integrated high-speed transceivers rapidly evolve. In national defense applications, the spectrum is further reduced with the presence of evolving adversary wideband communications systems. 

A common digital multi-carrier modulation technique that is a foundation of today’s wireless systems (e.g., 4G/5G, Wi-Fi, Digital TV, etc.) is also widely used in the adversary systems noted earlier. The technique, Orthogonal Frequency-Division Multiplexing (OFDM), is ubiquitous because it achieves high data rates using multiple lower-rate streams modulated in parallel, where each stream is modulated with a separate subcarrier. The subcarriers are orthogonal, which enables spectral overlap without inter-carrier interference (ICI). Having multiple narrowband subcarriers helps to mitigate fading. In addition, because the subcarriers are spaced very close together,  a large number can fit within an allocated channel bandwidth. The total bandwidth of an OFDM signal is roughly equal to the sum of the individual subcarrier bandwidths, which makes it much wider than a typical single-carrier signal of interest. The subcarrier modulation and spacing provide the structure in the time-frequency domains that an AI-enabled approach can learn via training data.

\subsection{Traditional Interference Rejection Techniques}

There are a range of classical techniques that can be used for interference rejection or mitigation. For radio systems where a phased array antenna is available on the RF front-end, the options include beamforming and adaptive nulling to reduce the effects of interfering signals. For the scenario in which this paper is focused, we assume only one channel is available at the receiver due to SWaP constraints. Given a single receive channel, our approach is limited to temporal and spectral processing on the mixed received signal (SOI plus interferer plus noise) and spatial separation is not considered.

Many traditional techniques for interference rejection where a strong OFDM interferer is present rely on signal processing techniques that focus on the properties of the SOI without considering the interferer. Since the time-frequency structure of an OFDM signal is known or can be learned, these types of traditional approaches do not provide the best solution. For example, a matched filter will maximize the signal-to-noise ratio (SNR) for a known signal in the presence of additive white Gaussian noise (AWGN) but it will not be optimal when the OFDM-based structured interference is present in addition to AWGN \cite{proakis2008digital}.

Alternatively, the Linear Minimum Mean Squared Error (LMMSE) technique will predict an unknown variable from noisy data by minimizing the average squared error \cite{kay1993fundamentals}. Unlike the matched filter, LMMSE exploits the interferer's correlation structure which results in better interference rejection performance. There are two drawbacks with the LMMSE approach. First, the calculations for the LMMSE involve computing covariance matrices where the eigenvalue decomposition step is slow and scales poorly with matrix size \cite{10945811}. These computations were timed on a server-grade NVIDIA H100 NVL GPU and found to be too slow for our application, where latency is critical (112811 msec for a signal length of 10240 baseband samples). In addition, one of the goals for the proposed technique is to handle emerging interference sources (including SINR levels) that are not fully known, understood, or characterized in detail. These factors combine to make the LMMSE method impractical.

Successive Interference Cancellation (SIC) multi-user detectors are also effective when the interference source is known. This technique operates by sequentially decoding the strongest received signal. The signal is then reconstructed and then subtracted to remove it from the original signal. This process is repeated until the SOI is decoded. There are challenges with this approach as well in that detailed knowledge of the channel and interference source will not be characterized \cite{moshavi1996multiuser}.

\section{Approach}

\subsection{AI-based Algorithms}

AI-enhanced interference rejection techniques can be employed to account for both the SOI and the interfering source. These techniques can also be applied to scenarios where limited information is known about the interfering signal.

\subsubsection{RF WaveNet}

\begin{figure}[htbp]
\centerline{\includegraphics[width=0.75\columnwidth]{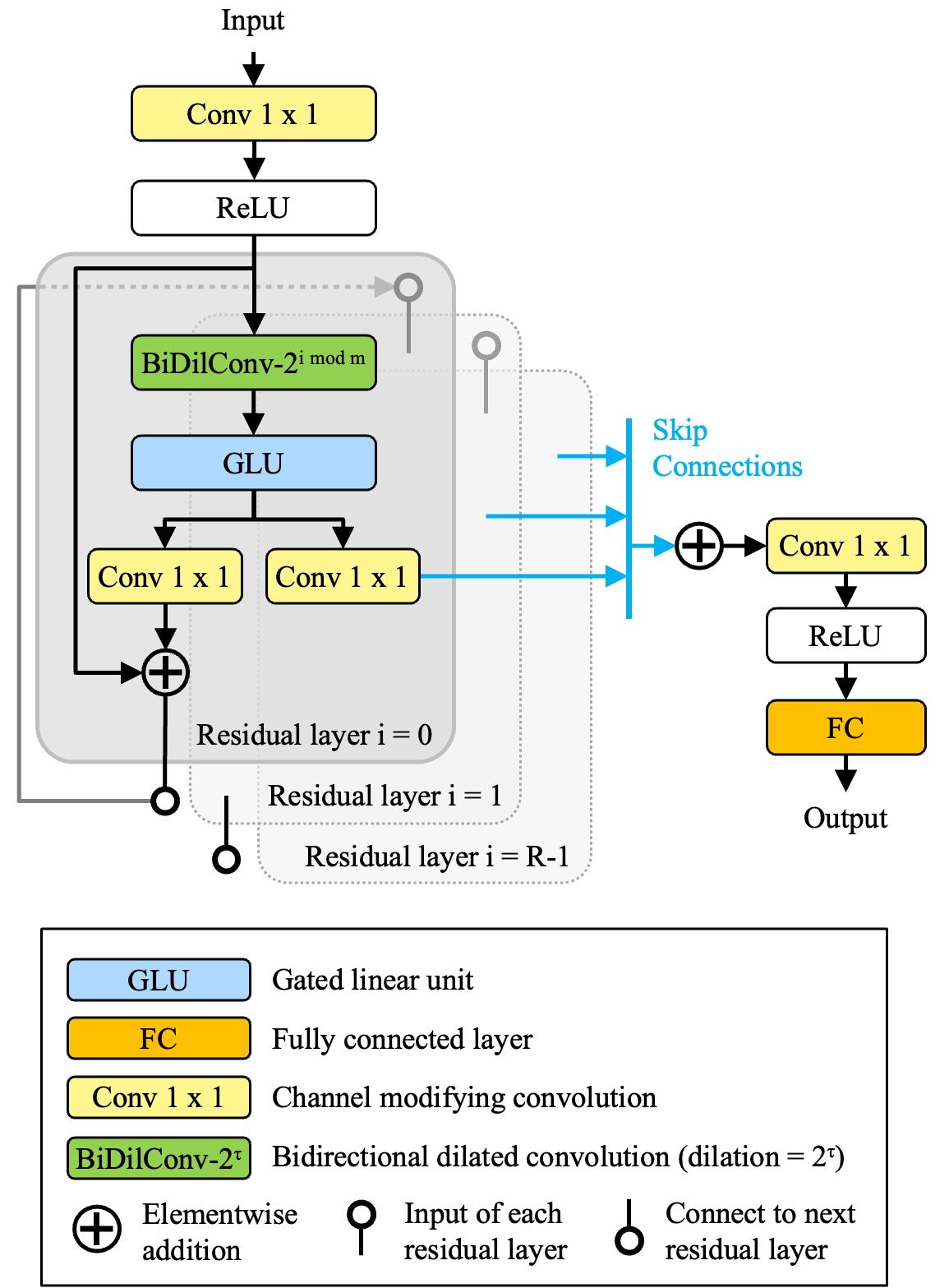}}
\caption{RF WaveNet, the WaveNet model adapted for RF data \cite{10945811, 45774}}
\label{fig:wavenet}
\end{figure}

The RF WaveNet architecture \cite{10945811} builds upon the original WaveNet model \cite{45774}, which achieved notable success in audio signal generation and modeling. RF WaveNet adapts this architecture for RF signal separation, leveraging its ability to capture long-range temporal dependencies through dilated convolutions, residual layers, and skip connections, as illustrated in Fig. \ref{fig:wavenet}. Unlike downsampling-based architectures such as U-Net, WaveNet preserves temporal resolution while expanding the receptive field using exponentially increasing dilation rates within residual blocks. For RF-specific tasks, the model processes complex-valued signals represented as I/Q components concatenated into separate channels, employs mean squared error (MSE) loss for training, and applies random time shifts and phase rotations to simulate transmission impairments. The real and imaginary components of our signal mixtures are separated, stacked, and passed to the model. These adaptations, along with an increased channel dimension $C=128$ and optimized dilation cycle, enable RF WaveNet to effectively separate signals-of-interest from interference.

\subsubsection{RF Transformer \& RF Transformer Decoder}

\begin{figure}[htbp]
\centerline{\includegraphics[width=0.75\columnwidth]{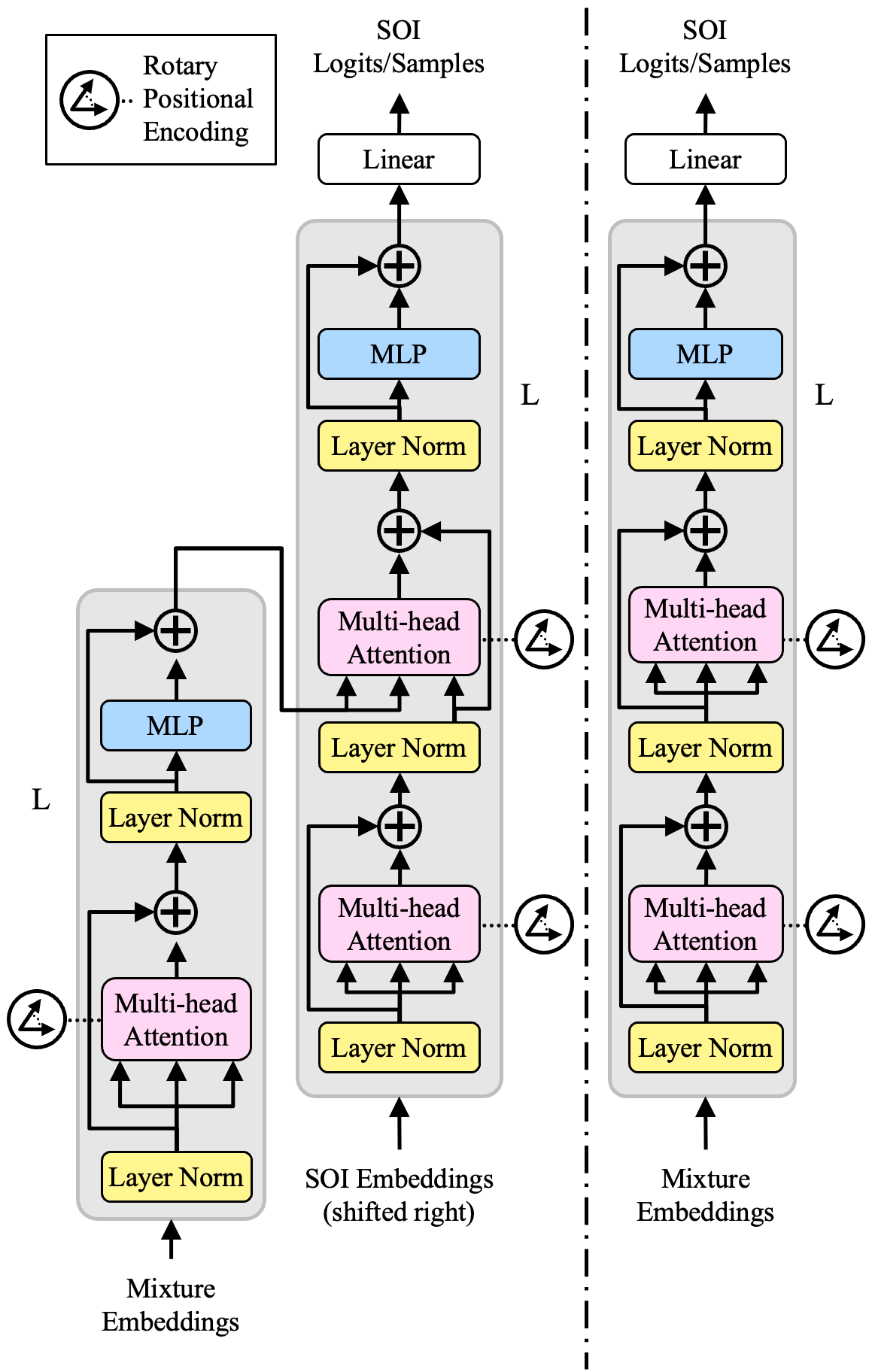}}
\caption{RF Transformer (left) \& RF Transformer Decoder (right), the Transformer models adapted for RF data \cite{jayashankar2025, LifarRFTransformer, 10.5555/3295222.3295349}}
\label{fig:transformers}
\end{figure}

The RF Transformer architecture \cite{jayashankar2025, LifarRFTransformer} builds upon the Transformer model \cite{10.5555/3295222.3295349} by adapting it for the RF domain. While the U-Net and WaveNet models process entire waveforms in parallel and require buffering larger segments of data, the RF transformer operates autoregressively and processes arriving RF signal samples sequentially. As illustrated on the left of Fig. \ref{fig:transformers}, the incoming mixture waveform is projected into a higher dimensional embedding space using a linear transformation and the causal attention mechanism ensures that each segment is only dependent on current and past segments. Additionally, the attention mechanism operates with a key-value cache to reuse attention computations, while rotary positional encodings improve the model's ability to capture temporal dependencies. The decoder portion generates signal of interest samples using self-attention and cross-attention to combine information from the mixture embeddings and previously decoded signal of interest embeddings. The transformer decoder can also operate independently, focusing solely on autoregressive generation of the signal of interest by directly attending to the mixture embeddings without requiring additional conditioning from the encoder portion. This makes the decoder ideal for streaming scenarios where real-time processing is required as shown on the right of Fig. \ref{fig:transformers}. Similar to RF WaveNet, during training the real and imaginary components are stacked, Mean Squared Error (MSE) is used as the loss function, and random time shifts and phase rotations to simulate transmission impairments are applied.

\subsection{Training Methodology}\label{method}

In order to train our models on RF signals, careful preprocessing was applied so that the signal of interest can be mixed with the interference at varying levels of SINR and remain representative with how RF mixtures would be formed over the air. 

\subsubsection{Data Preprocessing}

\begin{figure*}[htbp]
\centerline{\includegraphics[width=0.8\textwidth]{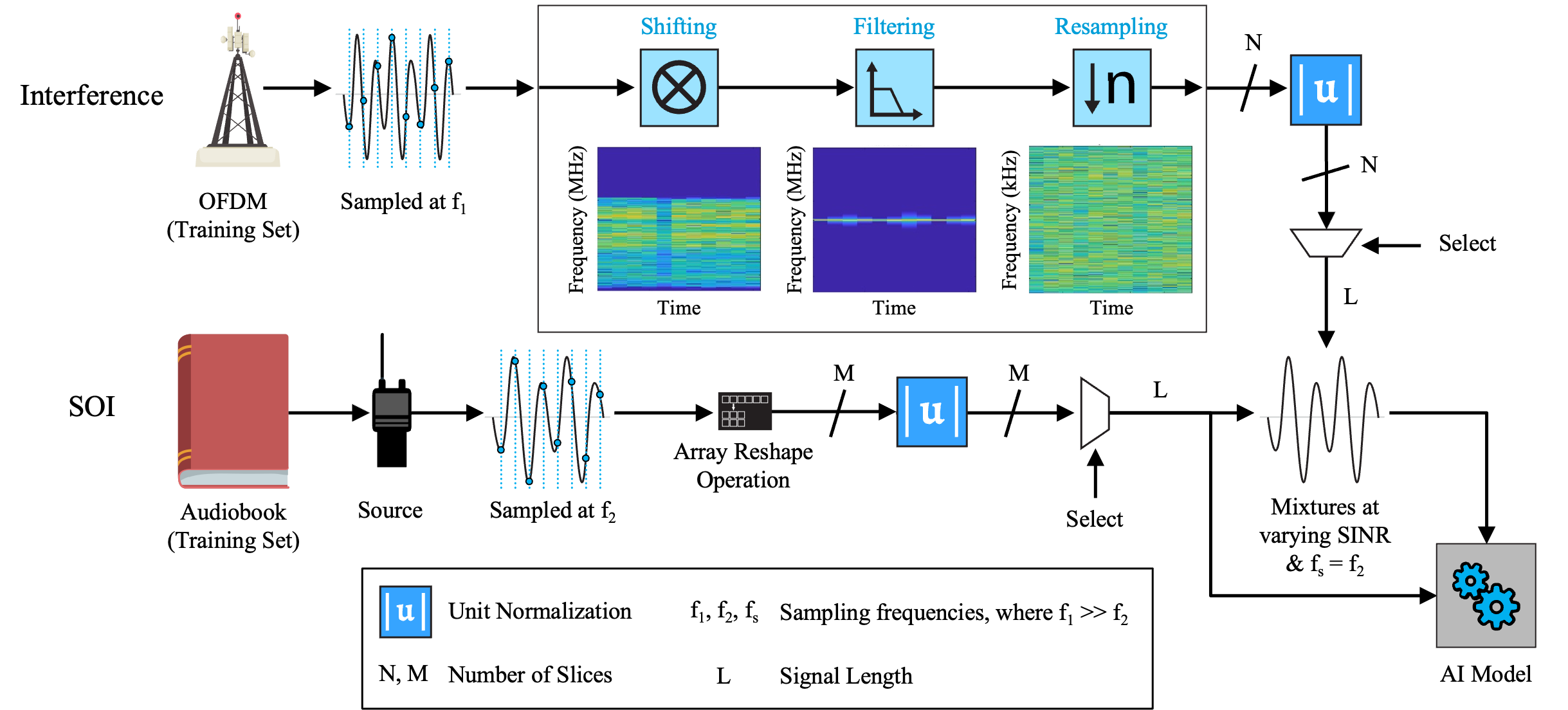}}
\caption{Data preparation process for the training pipeline. The interference is sampled at $f_1$, then shifted, filtered, and resampled, and then each of the N slices is unit normalized. The SOI is sampled at $f_2$ and then reshaped and unit normalized. A slice of interference and a slice of SOI are then selected at random and mixed at varying SINR levels and presented to an AI model for training.}
\label{fig:data_preprocessing}
\end{figure*}

Fig. \ref{fig:data_preprocessing} shows the full sequence of data preprocessing steps. Because the interference is a wideband signal, the signal of interest will not overlap with the entire band of interference in the frequency space. A series of transformations including a frequency shift, filter, and resample operation can be applied to let us reuse our interference data more efficiently and also allow different parts of the interference to be mixed with the signal of interest. The frequency shift cycles the interference through the frequency space, the filter removes everything except the interference content around baseband, and the resample operation brings the interference to a common sample rate. After applying these operations, the result is $N$ slices of length $L$ of interference, and each slice can be mixed with a slice of the signal of interest which can similarly be reshaped into $M$ slices of length $L$. In addition, when signals are received over the air they may have larger or smaller magnitudes depending on the gain of the receiver, so during preprocessing the magnitudes of both signals are unit normalized to ensure consistency and comparability between sequences, regardless of variations in receiver gain or signal strength. This normalization step helps mitigate the effects of amplitude differences and allows the model to focus on the intrinsic characteristics of the signals rather than their absolute magnitudes. Additionally, unit normalization prevents large variations in magnitudes during training which will help to avoid exploding or vanishing gradients or numerical instabilities that would lead to a poorly trained model.

\subsubsection{SINR Definition}

The mixture of the SOI and interference slices happen at varying levels of Signal-to-Interference-plus-Noise Ratio (SINR) so that the AI models are exposed to a range of easier and more difficult examples. By doing this, the AI models learn to remove interference across the wide range of SINR so that during test time it can properly handle an unknown SINR mixture which is within a range that we may see operationally. SINR is defined as the power of the desired signal $\textbf{s}$ divided by the combined power of the interference and background noise $\textbf{b}$, and is expressed in decibels (dB). Eqn. \ref{eq:sinr} defines SINR with scaling factor $\kappa$ where for a target SINR level $\kappa^2 = 10^{(\text{SINR in dB}) / 10}$ \cite{10945811}. Additionally, in order to mix the two slices correctly at the right SINR, the power of the two components is computed by limiting it to the region where the SOI's narrowband frequency content is present.

\begin{equation}
\text{SINR} = \frac{1}{\kappa^2} \frac{\mathbb{E}[ \| \mathbf{s} \|^2 ]}{\mathbb{E}[ \| \mathbf{b} \|^2 ]}
\label{eq:sinr}
\end{equation}

\subsection{Experimental Design: Analog Radio}

For the analog radio experiment, mixtures of analog FM voice as the SOI and synthetic 5G signals as the interference were processed. The choice of analog FM voice signals and 5G OFDM downlink signals for our experiments was driven by their technical relevance, strategic importance, and practical considerations. Analog FM voice signals are widely used in tactical communications due to their simplicity and robustness, making them a critical signal type for interference rejection in mission-critical scenarios. Our analog FM voice dataset was collected by transmitting a 15 hour audiobook of Charles Dickens' \textit{Tale of Two Cities} through a Baofeng model DM-1701 dual-band radio and receiving the RF signals that were emitted at a 50kHz sample rate. For interference, 5G OFDM signals represent a ubiquitous and complex interference source in modern RF environments, providing a challenging test case for our techniques. Our 5G OFDM dataset was generated with a sample rate of 15.36kHz and had a full 5G resource grid so that there was persistent interference on which to train. After applying the preprocessing discussed in section \ref{method}, the data was presented to our models for learning.

\section{Results}\label{Results}

In this section, the experiments are evaluated in terms of the processed signal quality as well as overall latency. These are the two most important factors for a tactical communications application and both must be acceptable in order to achieve a practical system. This section reports results for models with configurations described in Table \ref{tab:algorithms}.

\begin{table*}[htbp]
\caption{AI-based Algorithms Architecture Configurations}
\begin{center}
\renewcommand{\arraystretch}{1.2} 
\setlength{\tabcolsep}{3pt} 
\begin{tabular}{|c|c|c|c|c|}
\hline
\textbf{Model Name} & \textbf{RF WaveNet} & \textbf{RF Transformer} & \textbf{RF Transformer Decoder} \\
\hline
\textbf{Parameter Count} & 3964674 & 217550848 & 38913760 \\
\hline
\textbf{Input/Output Sequence Length} & \multicolumn{3}{c|}{10240 I/Q Samples} \\
\hline
\textbf{Number of Layers} & 30 residual & 12 encoder, 14 decoder & 14 decoder \\
\hline
\textbf{Hidden Dimension} & 128 residual channels & 768 & 480 \\
\hline
\textbf{Attention Heads} & N/A & \multicolumn{2}{c|}{12} \\
\hline
\textbf{Receptive Field} & 6139 & \multicolumn{2}{c|}{N/A} \\
\hline
\textbf{Window/Context Size} & N/A & 128 and 32 & 80 and 20 \\
\hline
\end{tabular}
\end{center}
\label{tab:algorithms}
\end{table*}

\subsection{Signal Quality: Audio Quality \& Intelligibility}

The processed signal quality was measured by comparing the reconstructed audio with the ground truth audio and two kinds of metrics were considered. These include quality, which measures how a speaker produces an utterance, and intelligibility, which is how clear the speaker's words are. Quality measures computed include Perceptual Evaluation of Speech Quality (PESQ) \cite{941023, miao_wang_2022_6549559}, Signal to Distortion ratio (SDR) \cite{raffel2014mir_eval}, Log-Spectral Distance (LSD) \cite{mcfee_2025_15006942}, and Mel-Cepstral Distance (Mel-CD) \cite{Taubert_mel-cepstral-distance_2025}. PESQ is a standard measure which models human hearing to score speech quality, SDR measures the level of distortion in the signal which impacts perceived quality, LSD evaluates differences in spectral content which affects timbre and clarity, and Mel-CD measures perceptual differences in Mel-frequency cepstral coefficients, which are tied closely to human auditory perception. Intelligibility measures include Short-Time Objective Intelligibility (STOI) \cite{pystoi, 10.1109/TASLP.2016.2585878} and extended STOI (eSTOI) \cite{pystoi, 10.1109/TASLP.2016.2585878}. As their names suggest, both STOI and eSTOI focus on evaluating the clarity of speech and ability to convey information to a listener. The plots for each of these metrics will have shaded regions for established scales that signify good, fair, and poor performance. The thresholds for each of these regions were chosen to provide a practical and interpretable framework. For metrics with established scales (PESQ, STOI, and eSTOI), the thresholds were informed by widely accepted benchmarks. For metrics without standardized scales (SDR, LSD, Mel-CD), the thresholds were chosen based on domain-specific expectations regarding the impact of distortions and spectral differences. There are a number of other metrics such as Perceptual Objective Listening Quality Analysis (POLQA) \cite{itu2011p863} that were not evaluated due to a lack of availability of a widely accepted open-source implementation for the metric or because they require a different sample rate for the source audio to evaluate. Finally, general metrics such as Mean Squared Error (MSE), Cepstral Distance (CD), and Entropy are not included as they have no bearing on audio quality or intelligibility.

\begin{figure}[htbp]
\centerline{\includegraphics[width=\columnwidth]{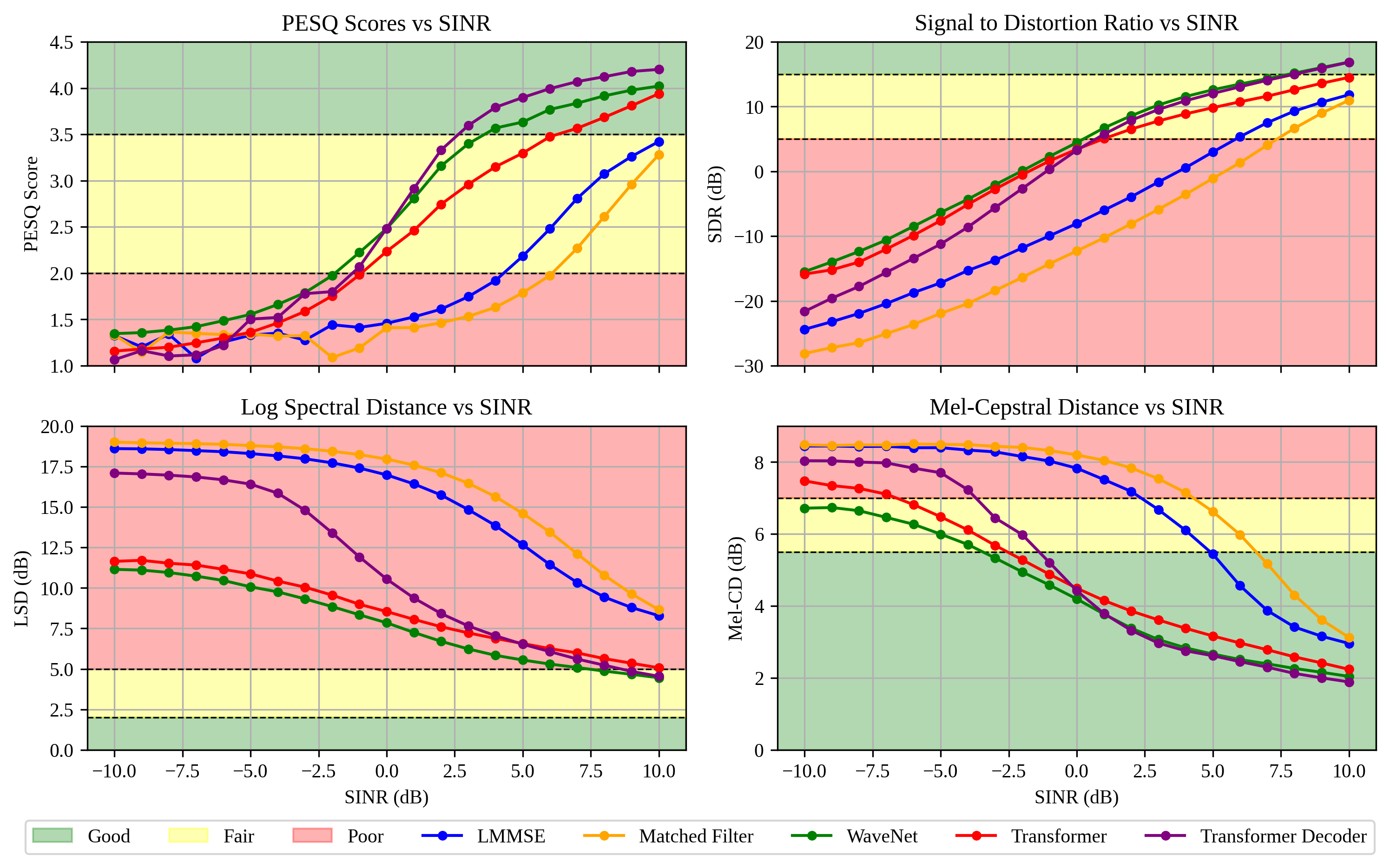}}
\caption{Audio quality evaluation for analog radio experiment. Metrics include Perceptual Evaluation of Speech Quality (PESQ) \cite{941023, miao_wang_2022_6549559}, Signal to Distortion ratio (SDR) \cite{raffel2014mir_eval}, Log-Spectral Distance (LSD) \cite{mcfee_2025_15006942}, and Mel-Cepstral Distance (Mel-CD) \cite{Taubert_mel-cepstral-distance_2025}.}
\label{fig:analog_audio_quality_vs_sinr}
\end{figure}

\begin{figure}[htbp]
\centerline{\includegraphics[width=\columnwidth]{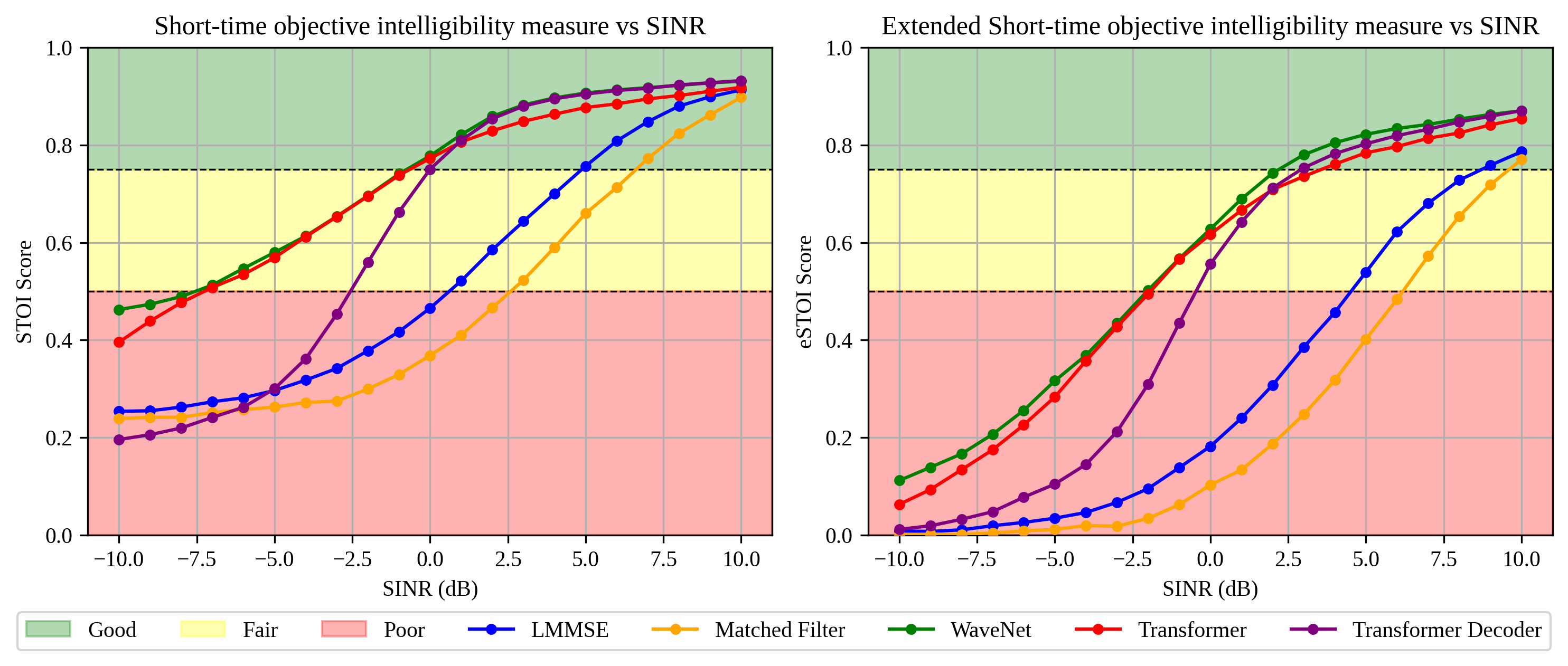}}
\caption{Audio intelligibility evaluation for analog radio experiment. Metrics include Short-Time Objective Intelligibility (STOI) \cite{pystoi, 10.1109/TASLP.2016.2585878} and extended STOI (eSTOI) \cite{pystoi, 10.1109/TASLP.2016.2585878}.}
\label{fig:analog_intelligibility_vs_sinr}
\end{figure}

For our analog radio experiment, audio reconstruction quality was measured using one minute of a held-out audiobook clip as the signal of interest and held-out 5G-compliant OFDM downlink signals. The two were mixed in the RF domain, passed through a model, and then converted back to audio using an FM demodulator as the matched filter. As seen in Fig. \ref{fig:analog_audio_quality_vs_sinr}, three out of the four metrics agree that audio quality is good or fair beyond SINR of 0 dB and the AI-based methods match or significantly out-perform the traditional methods. Fig. \ref{fig:analog_intelligibility_vs_sinr} outlines the results for the audio intelligibility metrics and both agree that the AI-based methods match or beat traditional methods. These results demonstrate how the AI-based methods perform well by explicitly considering characteristics of both the SOI and interference. This dual focus enables them to extract features from signal mixtures, leverage temporal and spectral context, and adapt to previously unseen interference sources.

\subsection{Overall latency}

The focus of this section is the two primary contributors to overall system latency: buffer latency and inference time. While additional latency may arise from factors such as sampling, transmission, data pipeline processing overheads, synchronization, and other hardware or network-related delays, these components are comparatively smaller and are not the focus of this discussion.

In a real-time streaming application, there is some delay introduced into the system by the amount of time it takes to accumulate enough samples to process through the model and this is known as the buffer latency. The buffer latency can be calculated as the number of samples that the model accepts for processing divided by the sample rate of the system. In the simplest case, the received samples can go unbatched where $B=1$. However, because many AI models benefit in terms of inference time from batching samples, the buffer latency will increase. In the case of higher sample rate data, this tradeoff may not pose as much of an issue. Additionally, the signal length that the AI model accepts is configurable and there are potential advantages to be gained from picking this parameter carefully. Eq. (\ref{eq:buffer_delay}) shows the relationship between batch size $B$, signal length $L$, and sample rate $f_s$ for buffer latency.

\begin{equation}
\text{Buffer Latency} = \frac{B \cdot L}{f_s}
\label{eq:buffer_delay}
\end{equation}

The inference time can be expressed by Eq. (\ref{eq:inference_time}) where it is a product of the forward pass time per window of signal length, $\tau(B, L)$, and the batch size $B$. The expression $\tau$ is dependent on the AI model's complexity (e.g., number of layers, operations), hardware acceleration (e.g., GPU, TPU, or CPU), and batch size and needs to be measured in practice. Note that inference time is not dependent on the sample rate, but a higher sample rate will require more inference forward passes and/or a higher batch size.

\begin{equation}
\begin{aligned}
\text{Inference Time} = \tau(B, L) \cdot B
\end{aligned}
\label{eq:inference_time}
\end{equation}

In order to have the system perform in real time, the overall latency must be kept to a minimum and the time to process a set of samples should be smaller than the time to buffer those samples. Additionally, there should not be a growing backlog of samples over a longer time window. In other words, the output throughput or the rate at which the system processes samples needs to be greater than or equal to the input throughput. The input throughput is simply the sample rate and the output throughput is defined in Eq. (\ref{eq:output_throughput}). Note that since we are operating in a streaming fashion, the buffering and inference actions are pipelined so the buffer time does not need to be included in the expression.

\begin{equation}
\begin{aligned}
\text{Output Throughput} \approx \frac{B \cdot L}{\text{Inference Time}}
\end{aligned}
\label{eq:output_throughput}
\end{equation}

\begin{figure}[htbp]
\centerline{\includegraphics[width=0.8\columnwidth]{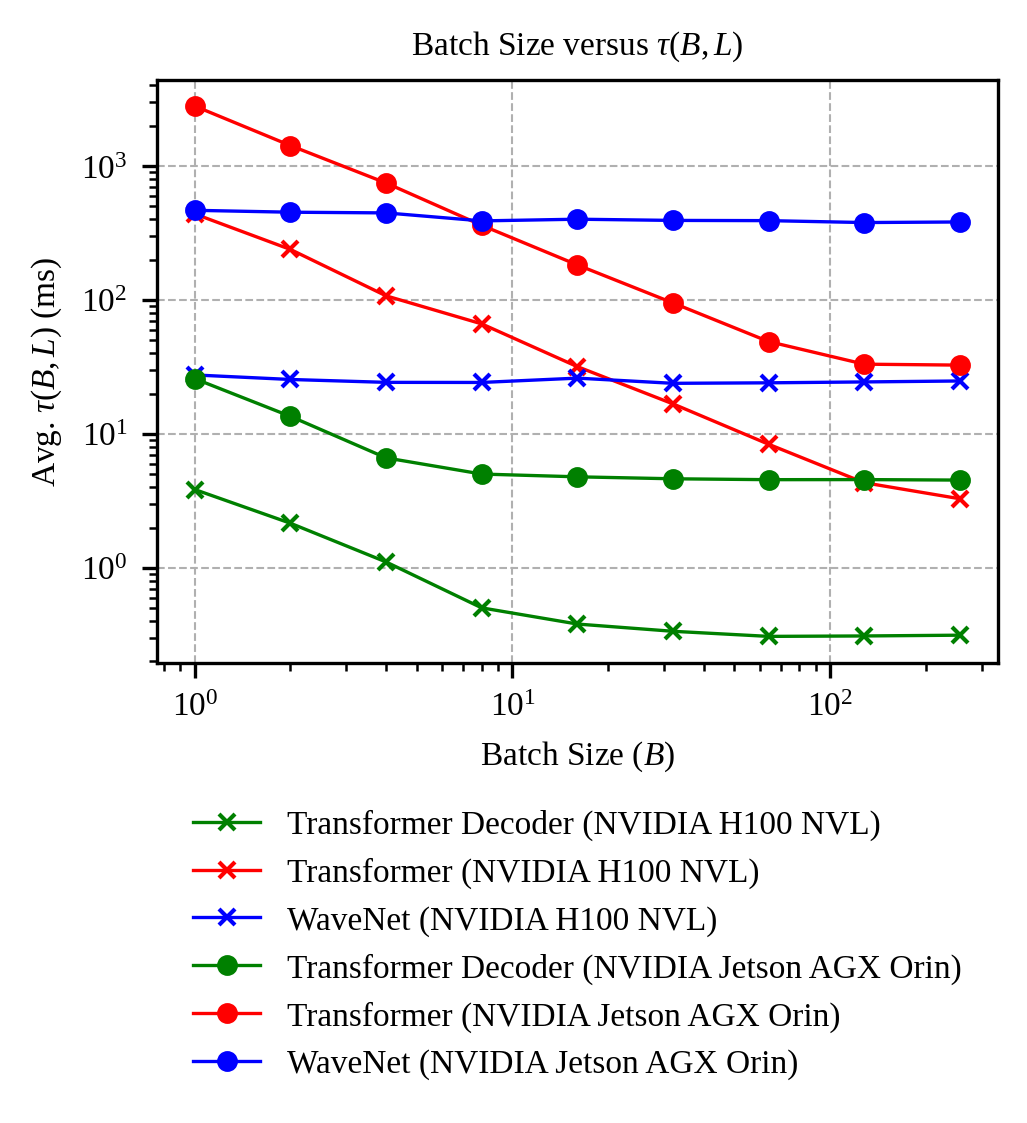}}
\caption{The impact of batching on inference time. As the batch size grows, the average  $\tau(B, L)$ decreases for transformer-based models which indicates the efficiencies for higher batch sizes. Note that this trend holds up to a certain point, beyond which memory constraints and diminishing returns can cause $\tau(B, L)$ to flatline or increase.}
\label{fig:batching}
\end{figure}

For the analog radio experiment, because the sample rate is only 50kHz, there is a heavy latency penalty associated with batching samples at the scale required to realize meaningful gains for inference time. For example, going from a batch size of 1 to 8, 16, or 32 would increase the buffer latency by a proportionate factor and cause the overall latency to balloon past real time. However, there are still gains to be had by tweaking signal length and in our findings, a size of 10240 I/Q samples worked well. For $B = 1$, $L = 10240$, and $f_s = 50$kHz, the buffer latency is estimated to be approximately 205 msec. Additionally for inference time, the $B = 1$ forward pass time was measured for our models on a NVIDIA Jetson AGX Orin edge GPU. Table \ref{tab:inference_times_analog} shows the results and includes a comparison with a server-grade GPU for reference. The overall latency for the transformer decoder model is estimated to be 205 msec + 25 msec $\approx$ 230 msec. We have also tested this in our lab on real hardware and observed about 10 msec of miscellaneous processing for an overall latency of 240 msec which is tolerable for a real-time streaming application. Additionally for $B=1$, the output throughput would be 10240 divided by 25 msec, or 410 kHz, which is much larger than the input throughput ($f_s = 50$kHz). Note that if we batch our samples, we can gain some inference time back but the cost will be a higher buffer latency. Fig. \ref{fig:batching} shows inference time in terms of forward pass time to process a window of 10240 I/Q samples versus various batch sizes. For datasets with higher sample rates, batching may be necessary to ensure that the output throughput exceeds the input throughput. This approach is advantageous because for transformer-based models, increasing the batch size typically reduces $\tau(B, L)$ due to improved computational efficiency and parallelization on the GPUs resulting in faster overall inference times. However, this trend holds only up to a certain point, beyond which memory constraints and diminishing returns can cause $\tau(B, L)$ to flatline or actually increase. Thus, while batching is beneficial, identifying an optimal $B$ is critical to balancing buffer latency, inference time, and throughput.

\begin{table}[htbp]
\caption{Inference Forward Pass Times $\tau(1, 10240)$}
\begin{center}
\renewcommand{\arraystretch}{1.2} 
\setlength{\tabcolsep}{3pt} 
\begin{tabular}{|c|c|c|c|c|}
\hline
\textbf{Hardware} & \textbf{RF WaveNet} & \textbf{RF Transformer} & \textbf{RF Transformer Decoder} \\
\hline
Server GPU$^{\mathrm{a}}$ & 28 msec & 439 msec & 4 msec \\
\hline
Edge GPU$^{\mathrm{b}}$ & 467 msec & 2905 msec & 25 msec \\
\hline
\end{tabular}
$^{\mathrm{a}}$ NVIDIA H100 NVL, $^{\mathrm{b}}$ NVIDIA Jetson AGX Orin\\
\end{center}
\label{tab:inference_times_analog}
\end{table}

\section{Discussion}

\subsection{Path to Deployment}

\begin{figure*}[htbp]
\centerline{\includegraphics[width=0.8\textwidth]{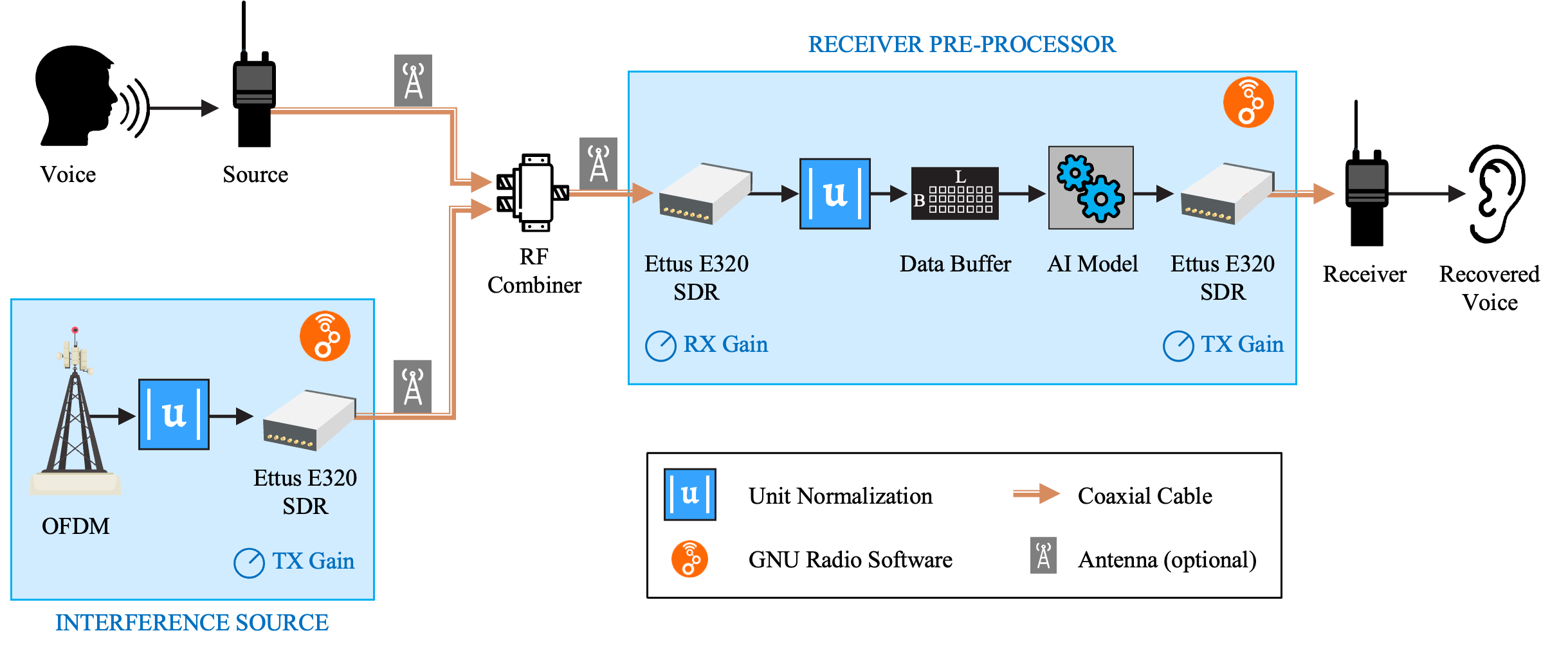}}
\caption{A conceptual setup for the real-time streaming application. The SOI and OFDM interference can be mixed either over the wire via the RF combiner or over the air (OTA) via antennas. Then the receiver will process the data by passing it through the AI model and then transmit it back to the receiver.}
\label{fig:streaming_setup}
\end{figure*}

Fig. \ref{fig:streaming_setup} shows a conceptual view of where an AI-enabled receiver preprocessor could be deployed at the tactical edge. Given the large number of tactical radios already in the field, a small receiver front-end could be added in place of existing radio antennas. This could be done without requiring any changes to the existing radios. Fig. \ref{fig:streaming_setup} shows the receiver preprocessor with two radios, but this is just for convenience, because the same radio chip can receive the signal and re-transmit it out to the antenna of a fielded radio.

In addition to the results shown in this paper, we have had initial success in using 5G-compliant downlink signals collected over the air (OTA) and used as the interference source, even though the network was only trained with synthesized interference data. This is promising given our goal of making the system robust in fully OTA scenarios.

\subsection{Future Work}

In the short term, our research will extend the transformer-based signal separation algorithm to digitally modulated RF signals and interferers. Beyond these new signals, our work will explore the adaptability of the AI-enabled algorithms to new OFDM-based interference sources. We will continue to compare our results with traditional techniques from a performance perspective.

Our work will also extend to expanded model robustness and generalization for over-the-air transmission. While all of the documented results described above have been achieved in lab space indoors, we plan to bring the systems to government-sponsored test ranges for further testing in outdoor propagation environments. 

Finally, we will optimize the AI-enabled architectures for better performance with direct applicability to higher sample rate datasets. This will open up the number of applications for which this technology can be used. Additionally, other architecture considerations will be explored to more efficiently handle higher sample rates. 

\section{Conclusion}

The results presented demonstrate considerable success in our goal of operationalizing recent results in AI-enhanced interference rejection to improve the performance of tactical radios. Fig. \ref{fig:analog_intelligibility_vs_sinr} exhibits an increase of roughly 7 dB SINR in the key metric of intelligibility (evaluated at the good-fair transition) for the transformer decoder vs. the matched filter baseline. In addition, Table \ref{tab:inference_times_analog} demonstrates that for transformer decoders, acceptable latency is attainable even with the hardware expected to be available under austere tactical conditions. Our Autoregressive Transformer Decoder models exhibit orders of magnitude faster throughput at inference time than both the WaveNet (feedforward DNNs) developed in earlier work and Transformers. 

In closing, we would like to broaden our view somewhat and add some final observations. First, we expect these results to generalize beyond the specific analog ``Walkie-Talkie'' radio case we investigated, although other use cases may favor somewhat different design choices. Second, our current use case is hardly the only one of interest for operationalizing AI-enhanced interference rejection. In general, by considering both SOI and interferer, AI-enhanced interference rejection in radio transmissions can be used to improve detection, demodulation, and decoding of signals over a range of SINR levels without having a detailed, design-level knowledge of the interfering signal or the propagation conditions. These same techniques can also be applied to a broader set of national security-related and commercial scenarios where interference rejection is a crucial issue.

\section*{Acknowledgment}
The authors acknowledge the MIT Lincoln Laboratory Supercomputing Center for providing HPC resources that have contributed to the research results reported within this paper \cite{8547629}. Additionally, we thank our MIT collaborators and the support given by Major Jovan Popovich, without whom this work would not have been possible.

\bibliographystyle{IEEEtran}
\bibliography{references}

\end{document}